\documentclass[preprint]{aastex}
\usepackage{emulateapj5}
\usepackage{graphicx}
\usepackage{rotate}
\usepackage{amsmath}
\usepackage{cancel}
\usepackage{amssymb}

\begin{document}

\title{Convective motions and net circular polarization in sunspot penumbrae}

\author{J.M.~Borrero$^{1}$ \& S.K.~Solanki$^{1,2}$}
\affil{$^{1}$Max Planck Institut f\"ur Sonnensystemforshung, Max Planck Strasse 2, Katlenburg-Lindau, 37191, Germany\\
$^{2}$School of Space Research, Kyung Hee University, Yongin, Gyeongg: 446-701, Korea}
\email{borrero@mps.mpg.de, solanki@mps.mpg.de}

\begin{abstract}
{We have employed a penumbral model, that includes the Evershed flow and convective motions
inside penumbral filaments, to reproduce the azimuthal variation of the net circular 
polarization (NCP) in sunspot penumbrae at different heliocentric angles for two 
different spectral lines. The theoretical net circular polarization
fits the observations as satisfactorily as penumbral models based on flux-tubes. The reason
for this is that the effect of convective motions on the NCP is very small compared to the effect
of the Evershed flow. In addition, the NCP generated by convective upflows cancels out the NCP generated
by the downflows. We have also found that, in order to fit the observed NCP, the strength of 
the magnetic field inside penumbral filaments must be very close to 1000 G. In particular, field-free or weak-field 
filaments fail to reproduce both the correct sign of the net circular polarization, as well as its dependence
on the azimuthal and heliocentric angles.}
\end{abstract}


\shorttitle{Convective Motions and NCP in Sunspot Penumbrae}
\shortauthors{BORRERO \& SOLANKI}
\maketitle

\def\nn{{\bf \nabla}}
\def\cro{\times}
\def\er{{\bf e_r}}
\def\et{{\bf e_{\theta}}}
\def\ex{{\bf e_x}}
\def\ez{{\bf e_z}}
\def\ey{{\bf e_y}}
\def\dep{(R,\theta)}
\def\depp{^{*}}
\def\deppp{(r,\theta)}
\def\depxz{(x,z)}
\def\psur{_{\rm gs}}
\def\pfil{_{\rm gf}}
\def\sur{_{\rm s}}
\def\fil{_{\rm f}}
\def\hx{\mathcal{H}}
\def\gx{\mathcal{G}}
\def\mx{\mathcal{M}}
\def\nx{\mathcal{N}}
\def\kms{~km s$^{-1}$}
\def\deg{^{\circ}}

\section{Introduction}%

Several investigations have proposed the presence of convective motions within the sunspot
penumbra (Danielson 1961, Grosser 1989, M\'arquez et al. 2006, Langhans 2006, S\'anchez Almeida 2005, 2006, 
S\'anchez Almeida et al. 2007), but only very recently have those motions been observationally pinpointed 
as occurring within penumbral filaments (Ichimoto et al. 2007; Rimmele 2008; Zakharov et al. 2008; cf. Bellot Rubio et al. 2005). 
Zakharov and co-workers have found that these convective flows appear similar to the
upper part of convective rolls proposed by Danielson (1961), with an upflow at the filament's 
center that turns into downflowing lanes at its edges. The measured speed of these motions is about $1$\kms.
Superposed to this convective flow is the Evershed flow, with typical
speeds of about $4-5$\kms, although much larger values
have been reported (del Toro Iniesta et al. 2001; Penn et al. 2003; Bellot Rubio et al. 2004;
Borrero et al. 2005; S\'anchez Almeida 2005). Recent 3D MHD simulations
(Scharmer et al. 2008a; Rempel et al. 2009) suggest a relation between 
these two velocity fields, with the Evershed flow being formed by the deflection
of the convective flow along the horizontal magnetic field inside penumbral filaments. \\

It is also well established that as the observer's line-of-sight penetrates
through the penumbral ambient field and into the penumbral filament, the magnetic inclination 
and line-of-sight velocity undergo large variations. These are widely
accepted as being responsible for creating the anomalous (i.e. asymmetric or even multi-lobed) 
polarization profiles observed in the penumbra (S\'anchez Almeida \& Lites
1992, Solanki \& Montavon 1993; see Solanki 2003 for a review). Models incorporating such variations
have successfully reproduced the azimuthal and Center-to-Limb variation of the 
net circular polarization (NCP) in visible and infrared Fe I lines
(S\'anchez Almeida 1996, 2005; Mart{\'\i}nez Pillet 2000, 2001; Schlichenmaier \&
Collados 2002; Schlichenmaier et al. 2002; M\"uller et al. 2002; 
Borrero et al. 2007). \\

However, the effect that the newly-discovered convective component
of the velocity field inside penumbral filaments has on the net circular
polarization (azimuthal and center-to-limb variation) has not
been studied. The convective flow can potentially have important 
consequences for the NCP observed at disk center, or at all disk positions
at locations perpendicular to  the line-of-symmetry of the sunspot. 
In both cases the Evershed flow is almost perpendicular to the line-of-sight, which should enhance the
contribution of the convective flow. Furthermore, the NCP generated by the convective flow
could be detected by spectropolarimeters operating at extremely
high spatial resolution (Scharmer et al. 2008b) and it could be
related to the non-zero NCP observed at the edges of
penumbral filaments (Ichimoto et al. 2008).

In this paper we address this possibility and study the effect
of the combined magnetic and convective flow pattern reported by Zakharov et al. inside 
penumbral filaments, on the azimuthal and center-to-limb variations
of the net circular polarization in sunspot penumbrae.

\section{MHS Model for penumbral filaments}%

We will adopt a 2.5D model for penumbral filaments similar to that of Scharmer 
\& Spruit (2006) and Borrero (2007). We assume that the properties of the 
filament do not change along its axis, i.e. directed radially outwards in the sunspot
($y$-axis). Therefore we can restrict ourselves to the XZ plane. In this plane the filament is located 
at the bottom of the domain: $z=0$. Hereafter we employ the indexes 'f' and 's' to refer to the filament
and its surroundings, respectively. The filament's boundary has a semi-circular shape of radius $R$. Using
polar coordinates ($r$,$\theta$), the magnetic field vector for the filament's interior, ${\bf B_f}$, 
and its surroundings ${\bf B_s}$ are prescribed as follows:

\begin{eqnarray}
  \begin{split}
    {\bf B_s}\deppp =  & B_0 \sin\gamma_0 \ey + B_0 \sin\theta\cos\gamma_0
    (1-\frac{R^2}{r^2})\er+ \\ & B_0 \cos\theta\cos\gamma_0
    (1+\frac{R^2}{r^2})\et \;\;\; \textrm{if $r > R$\;,}
  \end{split}
\end{eqnarray}

\begin{eqnarray}
{\bf B_f}\deppp = B_{\rm f0}\ey \;\;\;\; \textrm{if $r < R$\;,}
\end{eqnarray}

\noindent where $B_0$ and $\gamma_0$ refer to the strength and inclination (with respect
to the $z$-axis) of the surrounding magnetic field far away from the tube ($r \rightarrow \infty$). $B_{\rm f0}$
refers to the magnetic field inside the penumbral filament, which we assume to
be aligned with the filament's axis and homogeneous. We do not attempt to model what 
happens below $z = 0$ and therefore, throughout this paper, the polar angle coordinate 
is constrained to $\theta \in [0,\pi]$ (see Figure 1). Following Spruit \& Scharmer (2006) 
we have adopted a potential configuration for the surrounding magnetic field ${\bf B_s}$. 
Similarly, the velocity field is prescribed as follows:
  
\begin{eqnarray}
{\bf V_s} = 0 \;\;\;\; \textrm{if $r > R$\;,}
\end{eqnarray}

\begin{eqnarray}
  \begin{split}
    {\bf V_f}\deppp =  & V_e \ey + V_{\rm f r}(r) \er + \\ & V_{\rm f \theta}\deppp \et
\;\;\; \textrm{if $r < R$\;,}
  \end{split}
\end{eqnarray}

\noindent where $V_e$ refers to the Evershed flow (radial flow along the $y$-axis). Superposed to
it, we allow for the possibility of a convective flow pattern in the 
XZ plane. The radial, $V_{\rm fr}$, and angular $V_{\rm f\theta}$ components of the convective velocity 
flow are given by:

\begin{eqnarray}
V_{\rm fr}(r) = V_c \left\{1-e^{-\beta \left(r-R\right)^2}\right\}
\end{eqnarray}

\begin{eqnarray}
  \begin{split}
    V_{\rm f\theta}\deppp = & \bigg{[} \rho_0 \exp\{-r\sin\theta/H\sur\}+r \delta \sin\theta \bigg{]}^{-1} 
\bigg{\{}\frac{\partial(r V_r)}{\partial r} \cro \\ &
\left[\rho_0\left(\theta-\frac{\pi}{2}\right)+\alpha r \cos\theta\right] + r V_r \bigg{[}\alpha\cos\theta + \\ &
\frac{r\rho_0}{2 H\sur^2}\left(\theta-\frac{\pi}{2}-\cos\theta\sin\theta \right)\bigg{]}\bigg{\}}
\end{split}
\end{eqnarray}

\noindent where $\rho_0$ refers to the density in the surrounding atmosphere at $z=0$: $\rho_0=\rho\sur(0)$. 
In addition, $\alpha$ and $\delta$ can be written as:

\begin{eqnarray}
\alpha = & \frac{\rho_0}{H\sur}-\delta \;, \\
\delta = & \frac{B_0^2 \cos^2\gamma_0}{\pi g R^2} \;,
\end{eqnarray}

\noindent where $g$ represents the Sun's gravitational acceleration at the surface ($g=2.74\cro 10^4$ cm s$^{-2}$)
and $H\sur$ is the density scale height for the surrounding atmosphere in which the filament is embedded.
The value of $\beta$ in Eq. 5 can be chosen to allow for a more rapid/slow drop of the
radial (in the XZ plane) velocity profile within the penumbral filament. In our case we have chosen it
such that $\beta R^2 >>1$. This ensures that at $r=0$, $V_{fr}(0) \simeq V_c$. Thus $V_c$ can be identified 
with the magnitude of the convective upflow at the filament's center. The complicated  functional dependence 
of the velocity field comes from the fact that is has been derived
fully analytically under the following constraints: (a) Mass conservation inside the filament: 
$\nn(\rho_f {\rm \bf V}\fil)=0$; (b) Hydrostatic equilibrium inside the filament: $\nn P\pfil = \rho\fil {\bf g}$; 
(c) Total pressure balance between the filament and its magnetic surrounding; and (d) The overall 
configuration must be convective-like. The rather tedious derivation of ${\bf V}\fil\deppp$ is described 
in the Appendix of this paper. 

The resulting flow pattern inside the filament in the XZ plane is presented in Figure 1, where it can 
be seen that it features an upflow at the filament's center, with downflowing lanes at the 
filament's edges. This convective pattern resembles the flows inside penumbral filaments in the simulations
from Heinemann et al. (2007) and Rempel et al. (2009), as well as the pattern deduced from observations 
by Zakharov et al. (2008).

\begin{center}
\includegraphics[width=8cm]{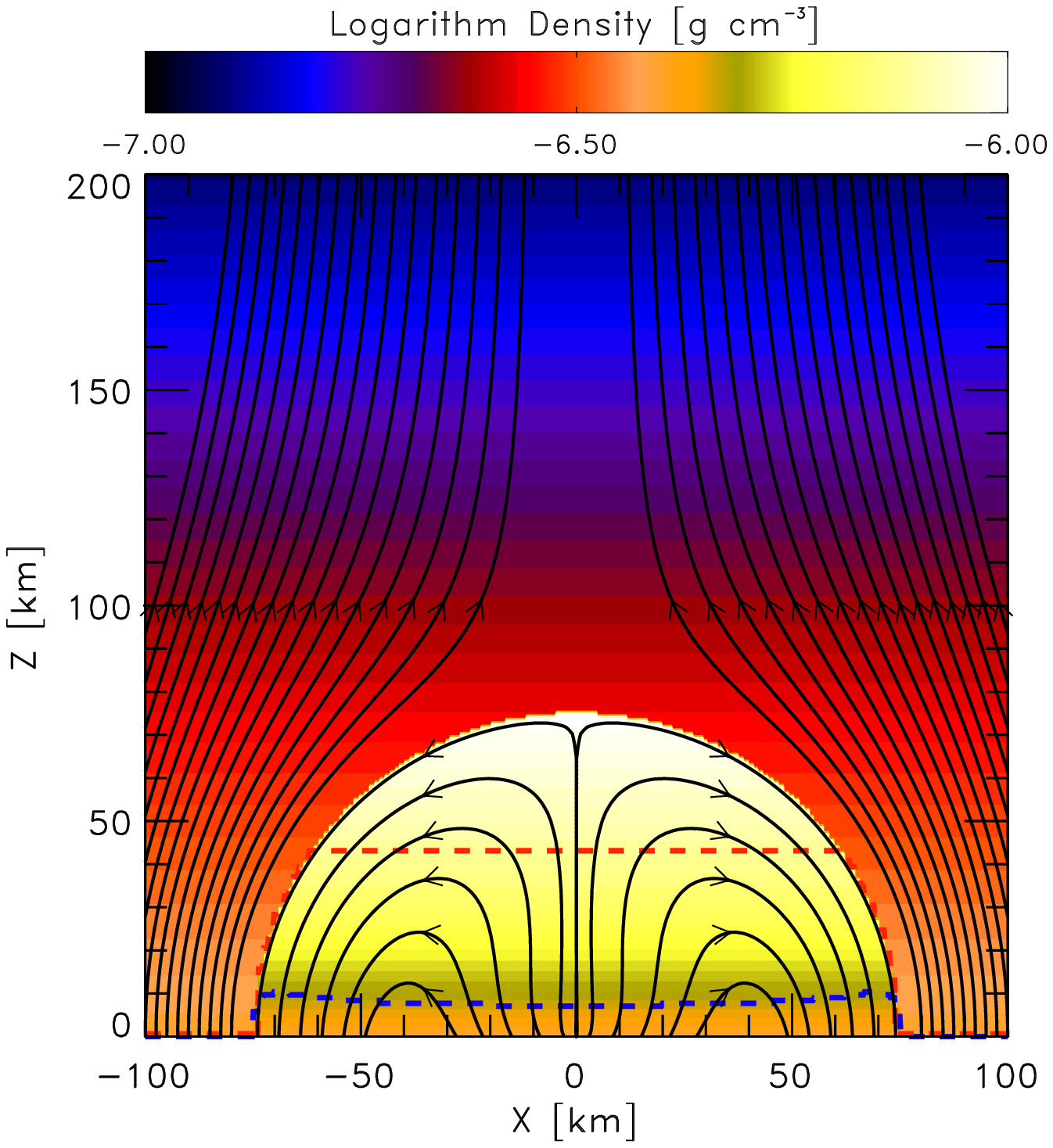}
\figcaption{Vertical cut (XZ plane) showing the density configuration for a penumbral filament in a 
surrounding potential field. This configuration was obtained with the following model parameters: 
$B_0 = B_{\rm f0} = 1000$ G, $R = 75$ Km, $\gamma_0 = 60\deg$. The field lines outside the filament 
($r > R$) correspond to the magnetic field lines in the surrounding atmosphere (Eq.~1) projected onto the XZ plane, 
while the field lines inside the filament denote the convective flow pattern (Eq.~4-8) in the same plane. The 
blue dashed line shows the location of the $\tau_5=1$ level (Wilson depression) and the dashed red line denotes
the location of the constant ($1.3\times 10^5$ dyn cm$^{-2}$) gas pressure level.}
\end{center}

Our model can be used to mimic the magnetostatic solutions
for the gappy penumbral model presented by Scharmer \& Spruit (2006). This can be achieved by 
simply reducing the magnetic field inside the filament until it becomes a field-free 
gap $B_{\rm f0}=0$. It can also mimic the classical flux-tube picture 
(Borrero 2007; Borrero et al. 2007) by removing the convective flow inside the 
filament $V_c=0$. The later two papers will be hereafter referred to as papers I and II.

Once the velocity and magnetic field have been prescribed, we can obtain the pressure
and density stratification of the surrounding atmosphere, $P\psur(z)$ and $\rho\sur(z)$, 
from a tabulated atmosphere. Here we use the hot umbral model from Collados 
et al. (1994; other models are discussed in Sect.~6). Note that, since the external magnetic field 
is potential, the adopted pressure and density stratification are also valid everywhere outside the 
filament. The boundary layer between the filament and the surrounding field is located at
$\dep \equiv (\pm\sqrt{R^2-z^2},z)$  (in polar and cartesian coordinates, respectively) and
denoted by a $\depp$ throughout this paper. At this boundary the following relation links the external and internal 
gas pressures:

\begin{eqnarray}
P\pfil\depp +\frac{B\fil^{*2}}{8\pi} = P\psur\depp+\frac{B\sur^{*2}}{8\pi}
\end{eqnarray}

This equation is valid irrespective of the external and internal velocity and magnetic fields, as long as
their radial components vanish at the filament's boundary: $V_{\rm fr}\depp=V_{\rm sr}\depp= B_{sr}\depp
=B_{fr}\depp=0$. As in paper I, we can take derivatives with respect to $\theta$ in Eq.~9 and apply the 
$\theta$-component of the momentum equation. This yields the following relation for the density
across the filament's boundary:

\begin{eqnarray}
\rho\fil\depp\left\{Rg\cos\theta+\frac{1}{2}\frac{\partial V_{\rm f \theta}^{*2}}{\partial \theta}
\right\}= \rho\sur\depp R g \cos\theta - \frac{1}{8\pi}\frac{\partial B_{\rm s\theta}^{2*}}{\partial \theta}
\end{eqnarray}

Note that the velocity we have prescribed (Eq.~6) satisfies $V_{\rm f\theta}\depp=0$. 
Therefore Eq.~10 can be simplified into:

\begin{eqnarray}
\rho\fil\depp = \rho\sur\depp+\frac{1}{\pi R g}\bigg{[}B_0^2 \sin \theta \cos^2\gamma_0\bigg{]}
\end{eqnarray}

These boundary conditions must be applied, together with the general stationary momentum equation, in order
to obtain the pressure and density structure inside the filament:

\begin{eqnarray}
\rho ({\bf v} \nn) {\bf v} & = & -\nn P_g + \frac{1}{c} {\bf j} \cro
{\bf B} + \rho {\bf g}
\end{eqnarray}

In this equation (ideal MHD without viscosity) the left-hand-side term corresponds to the advection term,
where the right-hand-side terms correspond to the gas pressure gradient, the Lorentz force and the
gravity, respectively. Since the magnetic field inside the filament is constant (Eq.~2), the Lozentz 
force (${\bf j} \cro {\bf B}$) plays no role in the pressure and density equilibrium for $r < R$. 
In addition, the convective velocities are much smaller than the sound speed: $V_{\rm fr}$, 
$V_{\rm f\theta} \simeq 1$\kms~ (see Ichimoto et al. 2007; Zakharov et al. 2008) and therefore
 the advection term can be neglected. This yields a pressure and density balance that conforms 
with hydrostatic equilibrium inside the filament: $\nn P\pfil = \rho\fil {\bf g}$. The horizontal
 ($x$-axis) component of this equation yields the pressure. Once it is obtained, the vertical 
($z$-axis) component of the momentum equation gives the density:
  
\begin{eqnarray}
\begin{split}
P\pfil(z) & = P\psur(z) + \frac{1}{8\pi}\Bigg{[} 4 B_0^2 \cos^2\gamma_0
\left(1-\frac{z^2}{R^2}\right) \\ & +  B_0^2 \sin^2\gamma_0 - B_{\rm f0}^2\Bigg{]}
\end{split}
\end{eqnarray}

\begin{eqnarray}
\rho\fil(z) = \rho\sur(z) + z \frac{B_0^2\cos^2\gamma_0}{\pi g R^2}
\end{eqnarray}

Figure 1 shows the density configuration for a penumbral filament and its surroundings, along
with the magnetic field lines outside the filament and the convective flow pattern inside it. 
Once the gas pressure and the densities are known, the temperature can
be evaluated using the ideal gas law with a variable molecular weight to account
for the ionization of the different species. As a result, we now have the temperature, density, 
gas pressure, and the velocity and magnetic field vector at every point in the XZ plane. 

Note that our approach to the magneto-hydrostatic 
equilibrium is slightly different from Scharmer \& Spruit (2006). This 
yields a different thermodynamic structure. For example, in Scharmer \& Spruit (2006) the density
inside the gap is larger than the density outside by a constant factor at all depths. In our case, 
the density difference changes linearly with depth (Eq.~14), and it peaks at 
the top of the filament while vanishing at $z = 0$. Our approach is also different from 
the flux-tube MHS equilibrium presented in Borrero (2007) in that we do not model the lower
half of the filament, as we do not know if deeper down the filament has the shape of a flux-tube or 
an elongated plume. This in turn means that we do not have to deal with possible negative densities
in the lower half of the filament as in the flux-tube case (see Eq.~14 in paper I). It also
allows us to have an uniform magnetic field inside the filament (Eq.~2). All these details about 
the thermodynamics, however, play a secondary role for radiative transfer 
calculations. In particular, they are negligible for the net circular polarization 
since this quantity depends mostly on the magnetic field and velocity configurations.

Finally it is also important to mention that Equations 2 and 4 ($\rm {\bf B_{\rm f}}\deppp$ and $\rm {\bf v_{\rm f}}\deppp$)
imply, through the induction equation, that the magnetic field along the filament's axis, $B_{\rm f0}$, changes in time.
This incovenience could have been avoided by postulating a magnetic field inside the filament
that is parallel to the velocity field. Since the magnitude of the Evershed effect is much larger than
that of the convective velocities: $V_{e} >> V_c$, it immediatelly follows that the magnetic field in the
XZ plane is much smaller than the magnetic field along the filament axis.

\section{Reference frame and azimuthal variation of the NCP }%

The thermodynamic, kinematic  and magnetic configuration for a penumbral filament has
 been obtained in the previous section in the Local Reference Frame: $\mathcal{S} = \{\ex,\ey,\ez\}$ 
(where the $z$-axis corresponds to the direction perpendicular to the solar surface), but in order
to study the azimuthal variation of the net circular polarization at different heliocentric angles
we need to place ourselves in the observer's reference frame: $\mathcal{S}^{''} =
\{\ex^{''},\ey^{''},\ez^{''}\}$, where now the $z^{''}$-axis points towards the observer.
To that end, we perform a double rotation of the velocity and magnetic field vectors.
First a rotation by angle $\Psi$ along $\ez$. This allows us to place the filament
at any azimuthal position within the sunspot. $\Psi = 0$ refers to the line-of-symmetry
of the sunspot on the center-side penumbra (i.e. it points towards the center of the solar 
disk). Secondly, a rotation by angle $\Theta$ (heliocentric angle) along 
the resulting $\ey^{'}$. This allows us to locate the 
sunspot at any position on the solar disk (see Eq.~1 in paper II). 
After performing these rotations, the inclination of the magnetic
field vector with respect to the observer can be obtained as: $\gamma = \cos^{-1}(B_z^{''}/B)$, 
the azimuth of the magnetic field in the plane perpendicular to the observer as:
$\phi = \tan^{-1}(B_x^{''}/B_y^{''})$, and finally the line-of-sight velocity as:
$v_{los} = v_z^{''}$.

The equations describing the ray paths (lines-of-sight) along which the radiative transfer equation is to 
be solved, is given by:

\begin{eqnarray}
x & = & x_0 + (z_{\rm max} - z)\tan\Theta\sin\Psi \\
y & = & y_0 + (z_{\rm max} - z)\tan\Theta\cos\Psi \textrm{\;,}
\end{eqnarray}

\noindent where $(x_0,y_0,z_{\rm max})$ is the point where the line-of-sight pierces the uppermost
boundary of our computational domain. Note that our model is 2.5D which means there are no variations
along the filament's axis ($y$-coordinate) which implies that Equation's 16 role can be simply subtituted by a
modification in the optical depth scale as $d\tau_{\rm los} = d\tau /\cos\beta$, with $\beta=\tan^{-1}(\tan\Theta\cos\Psi)$. 
With this information, we can now calculate the paths of each line-of-sight piercing the XZ plane in Figure 1 at different $x_0$'s. 
In our calculations we use 64 ray-paths with $x_0=-2R,...,2R$ . The radiative transfer equation 
is solved using the synthetis module in the SIR code (Ruiz Cobo \& del Toro Iniesta 1992) for each ray path.
Stokes $V$ profiles as a function of wavelength of two widely used spectral lines: 
Fe I 6302.5 \AA~ and Fe I 15648.5 \AA~ are computed. The net circular polarization for each ray, 
${\rm N}_{m}$ is obtained as the wavelength integral of Stokes $V$, with the final NCP (denoted as $\mathcal{N}$) 
being the mean over the ray-paths that pierce the filament (only $M$ rays out of 64)\footnote{By averaging
only over the lines-of-sight that pierce the filament we are ensuring that the filling factor of the
filament is always one or, in other words, that our resolution element is fully occupied by the filament
irrespective of $\Theta$ and $\Psi$. Failing to do this would allow us to arbitrarily change the filling factor
at each azimuthal position to create more or less net circular polarization. Note that the same results would be
obtained if we assume that there are several filaments lying next to each other within the resolution elements
as long as it is fully filled with filaments and they at located at the same height.}

\begin{equation}
\mathcal{N} = \frac{1}{M}\sum_{\displaystyle m}^{\displaystyle M} {\rm N}_m = \frac{1}{M}
\sum_{\displaystyle m}^{\displaystyle M} \int V_m(\lambda) d\lambda
\end{equation}

Figure 2 (top panel) shows examples (dashed lines) of the indiviudal 64 Stokes $V_m$ profiles generated
by each of the ray paths when looking at a penumbral filament located along the line-of-symmetry in the 
limbward-side of the penumbra ($\Psi=\pi$) of a sunspot located at an heliocentric angle of $\Theta=45^{\circ}$.
Only half of the ray-paths actually pierce the filament and produce a non-vanishing NCP ($M$ out of 64). In color 
we also plot the averaged Stokes $V$ profiles. The lower panel of Fig.~2 shows an example of the NCP generated 
by individual ray-paths (N$_m$) in the two considered spectral lines, at an heliocentric angle of $\Theta=45\deg$, 
and at two different azimuthal angles: $\Psi=0$ (center side penumbra) and $\Psi=\pi$ (the limb-ward side
of the penumbra). This example was obtained using the following model parameters: $B_0=B_{\rm f0}=1000$ G,
$V_e = 6$\kms, $V_c = 1$\kms, $\gamma_0 = 60\deg$ and $R = 75$ km.

\begin{center}
\includegraphics[width=8cm]{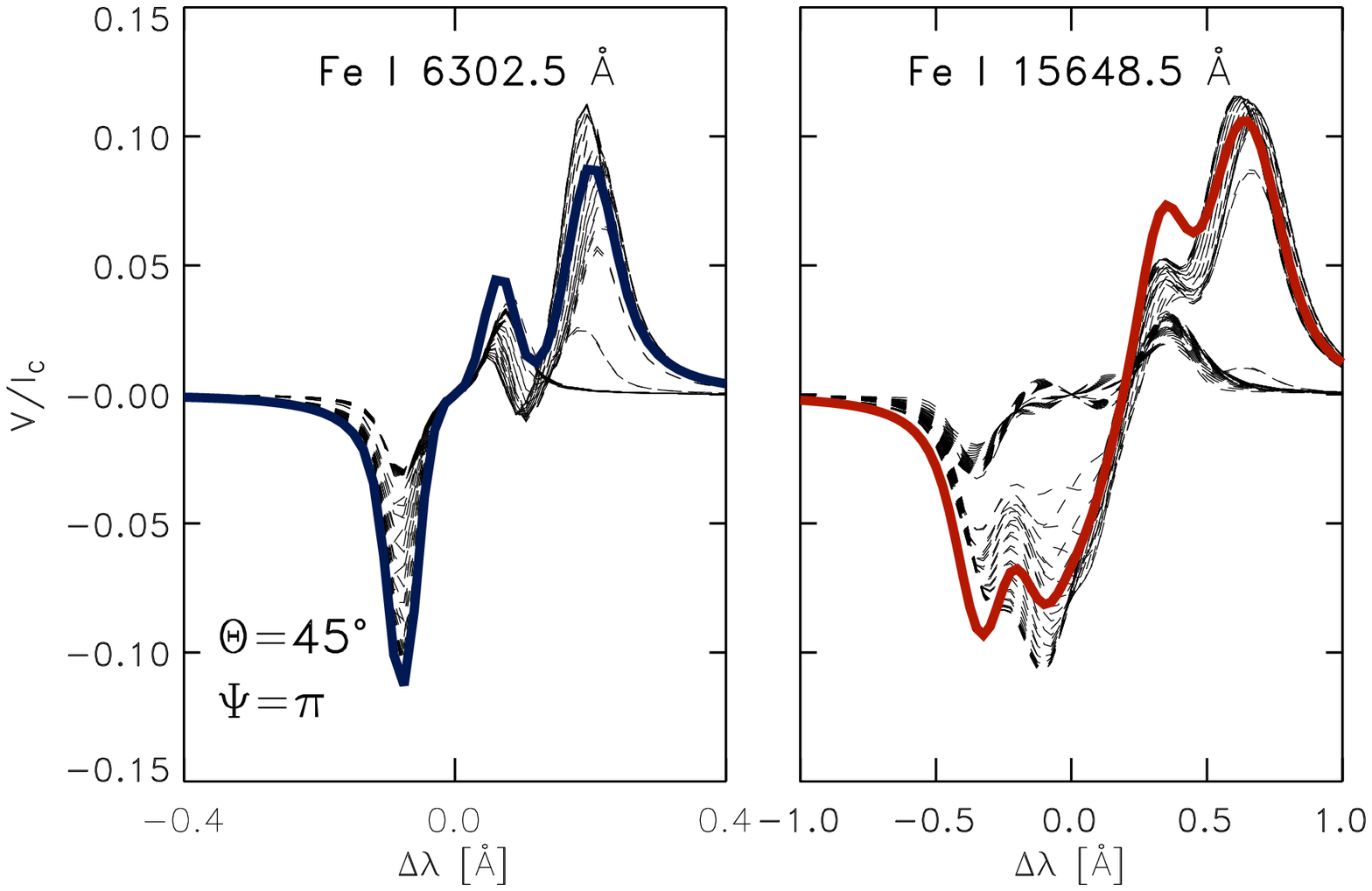} \\
\hspace{0.25cm}\includegraphics[width=8cm]{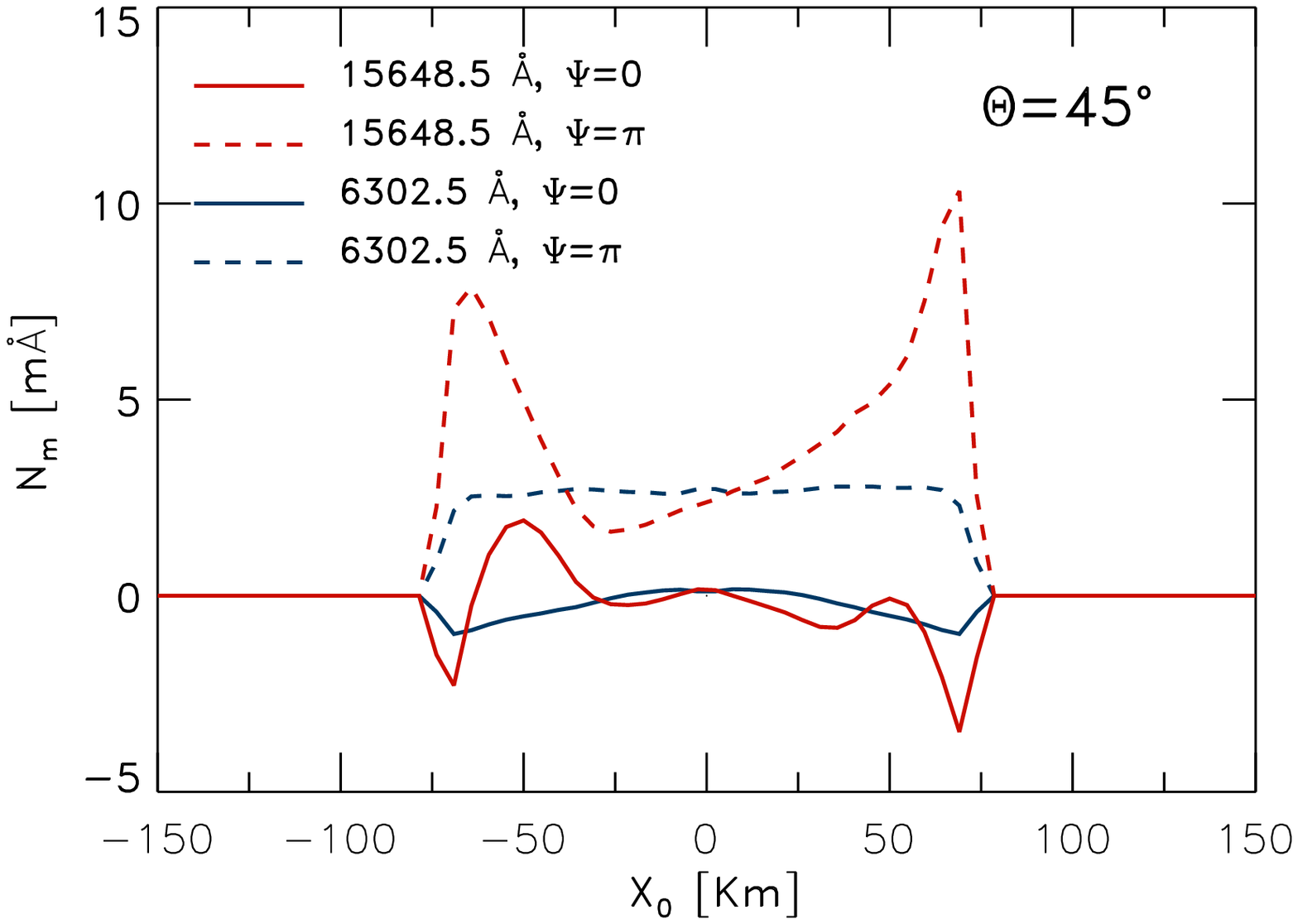}
\figcaption{{\it Top panel}: individual Stokes $V_m$ profiles (black dashed lines) generated by each 
of the ray-paths for $\Theta=45^{\circ}$ and $\Psi=\pi$. The averaged profile (over all 64 individual rays)
is also showed in color. {\it Bottom panel}: net circular polarization generated by different ray-paths,
 N$_m$, piercing the filament  at different points. Solid lines correspond to the center side penumbra on the line of
symmetry of the sunspot ($\Psi=0$), while dashed lines correspond to the limb side penumbra
also over the line of symmetry ($\Psi=\pi$). Blue lines correspond to the Fe I line at 6302.5 \AA~
and red lines are for Fe I 15648.5 \AA~. Model parameters are the same as in Figure 1: $B_0 = B_{\rm f0} = 
1000$ G, $R = 75$ Km, $\gamma_0 = 60\deg$. In addition, we have employed: $V_e = 6$\kms, $V_c = 1$\kms.}
\end{center}

We have repeated the same calculations for 25 different azimuthal positions between $\Psi = 0,2\pi$
and at 4 heliocentric angles: $\Theta = 15, 30, 45, 60\deg$. Results
are presented in Figure 3 (top panel for Fe I 6302.5 \AA~ and bottom panel for Fe I 15648.5 \AA~).
Example of theoretical and observed  $\mathcal{N}(\Psi)$-curves are overplotted in Figure 4
for two cases: ASP (Advanced Stokes Polarimeter; Elmore et al. 1992) observations of Fe I 6302.5 \AA~ at $\Theta=38\deg$
(AR 8545; May 21, 1999) and TIP (Tenerife Infrared Polarimeter; Mart{\'\i}nez Pillet et al. 1999) 
observations of Fe I 15648.5 \AA~ at $\Theta=60\deg$ (AR 8706, September 21, 1999). Figure 4 
clearly demonstrates that the total amount of NCP and its sign are well reproduced as a function
 of the azimuthal position at the displayed heliocentric angles. It is particularly gratifying to
 see the model reproducing the multi-peak shape of the NCP curve of Fe I
15648.5 \AA. The theoretical $\mathcal{N}(\Psi)$ curves compare satisfactorily with the observed ones
also at other heliocentric angles for these two spectral lines (compare Fig.~3 with Figs.~3-4 of 
paper II). 

\begin{center}
\includegraphics[width=8cm]{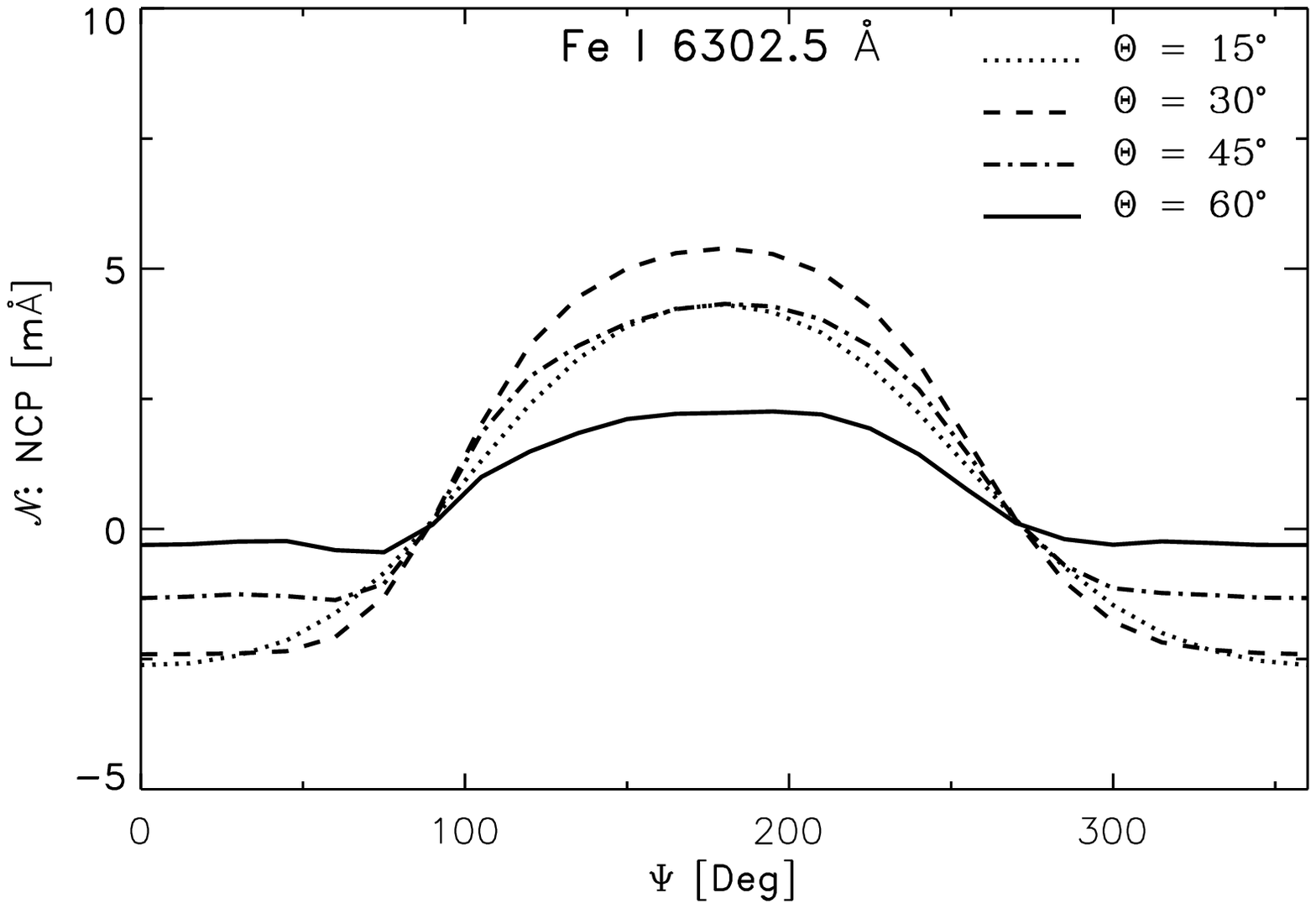} \\
\includegraphics[width=8cm]{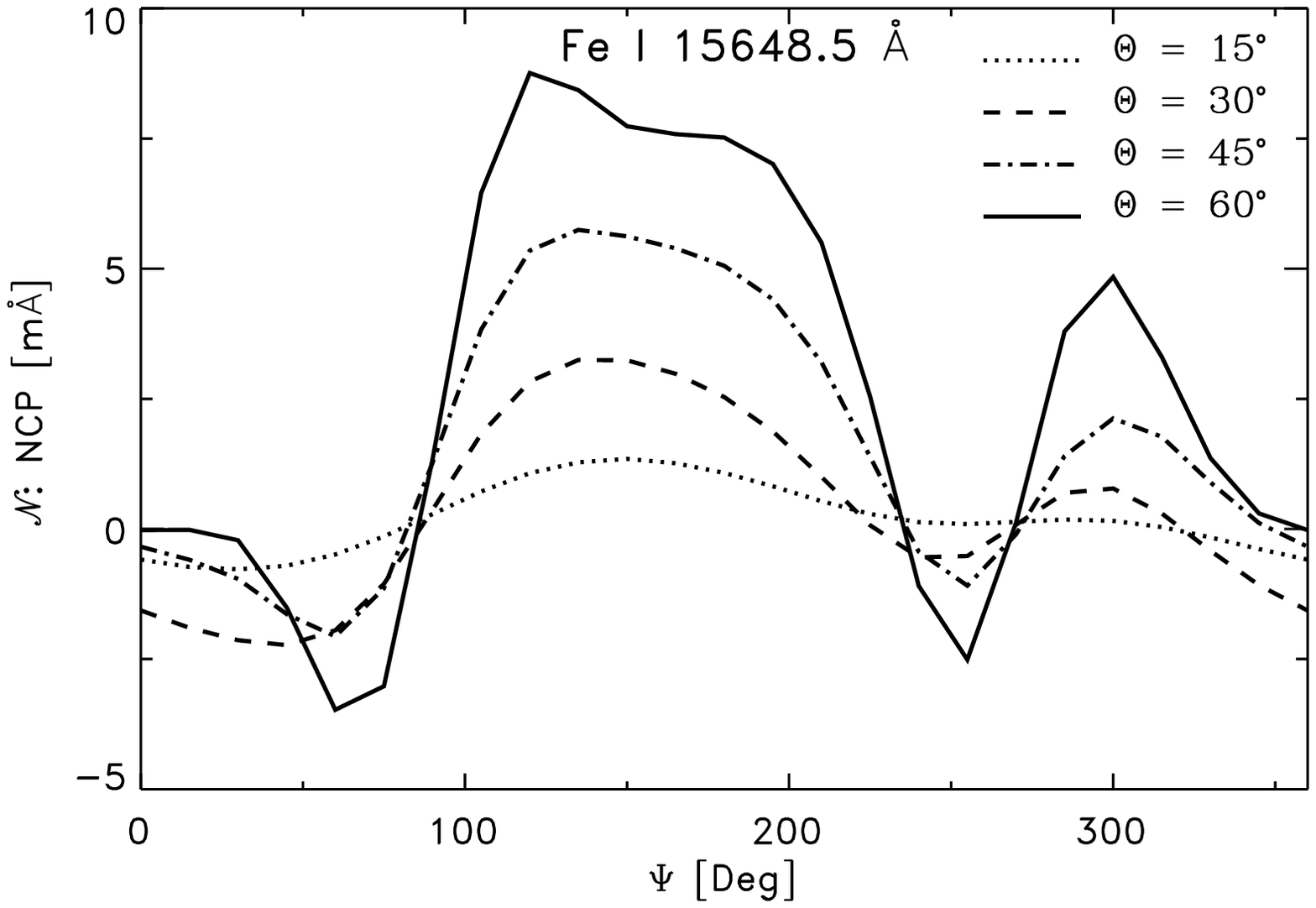}
\figcaption{Predicted azimuthal variation of the net circular polarization, $\mathcal{N}(\Psi)$,
for two different neutral iron atomic lines: 6302.5 \AA~ (top) and 15648.5 \AA~ (bottom)
for sunspots located at different heliocentric angles. They have been obtained
using the model for penumbral filaments described in this paper, which includes both the
Evershed flow and convective motions inside the filament. The model parameters used are the same 
as in Figure 2: $B_0 = B_{\rm f0} = 1000$ G, $R = 75$ Km, $\gamma_0 = 60\deg$, $V_e = 6$\kms, $V_c = 1$\kms.}
\end{center}

It is important to mention here that the observed NCP curves have been obtained mainly for points located
in the middle penumbra. It may seem that the model parameters: $B_0=B_{\rm f0}=1000$ G
and $\gamma_0 = 60\deg$ chosen to reproduce them are more typical of the outer penumbra. This is not
the case since these model parameters refer to locations far away from the flux-tube. In fact, in the
vicinity of the flux-tube the field strength and inclination of the external magnetic field
reaches values closer to 1500 G and 45$\deg$ respectively (see for example Fig.~1 in paper 1),
which is more representative of the mid-penumbra.

\section{Effect of a convective flow on the NCP}%

As demonstrated in the previous section, the model for penumbral filaments employed here
produces very similar $\mathcal{N}(\Psi)$-curves as the round horizontal flux-tube
model employed by Borrero et al. 2007 (see Figs.~3-4) to describe penumbral filaments.
In order to understand the reason for this we need to investigate the similarities and 
differences between our current model for penumbral filaments and the model from paper II.

\begin{center}
\includegraphics[width=8cm]{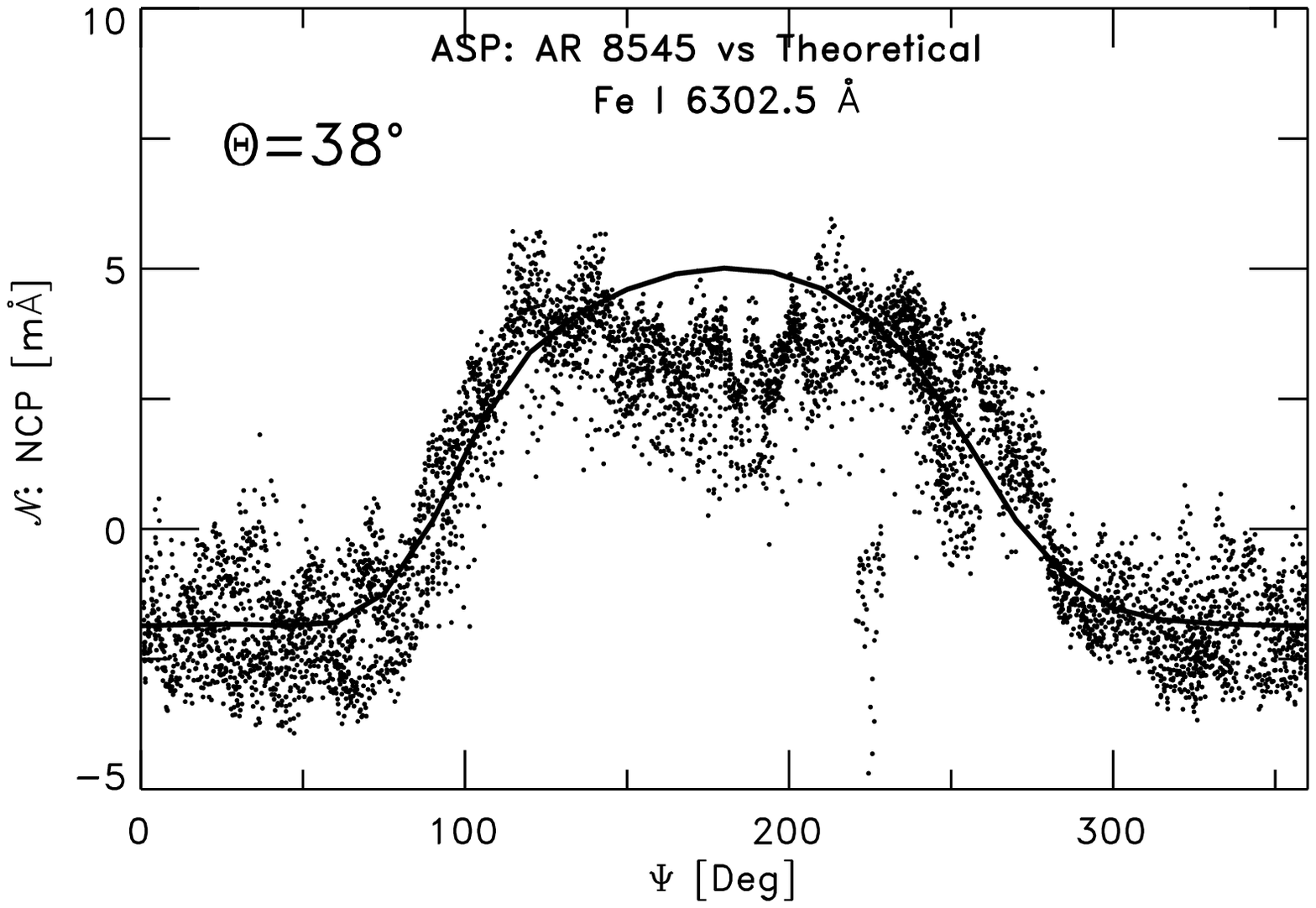} \\
\includegraphics[width=8cm]{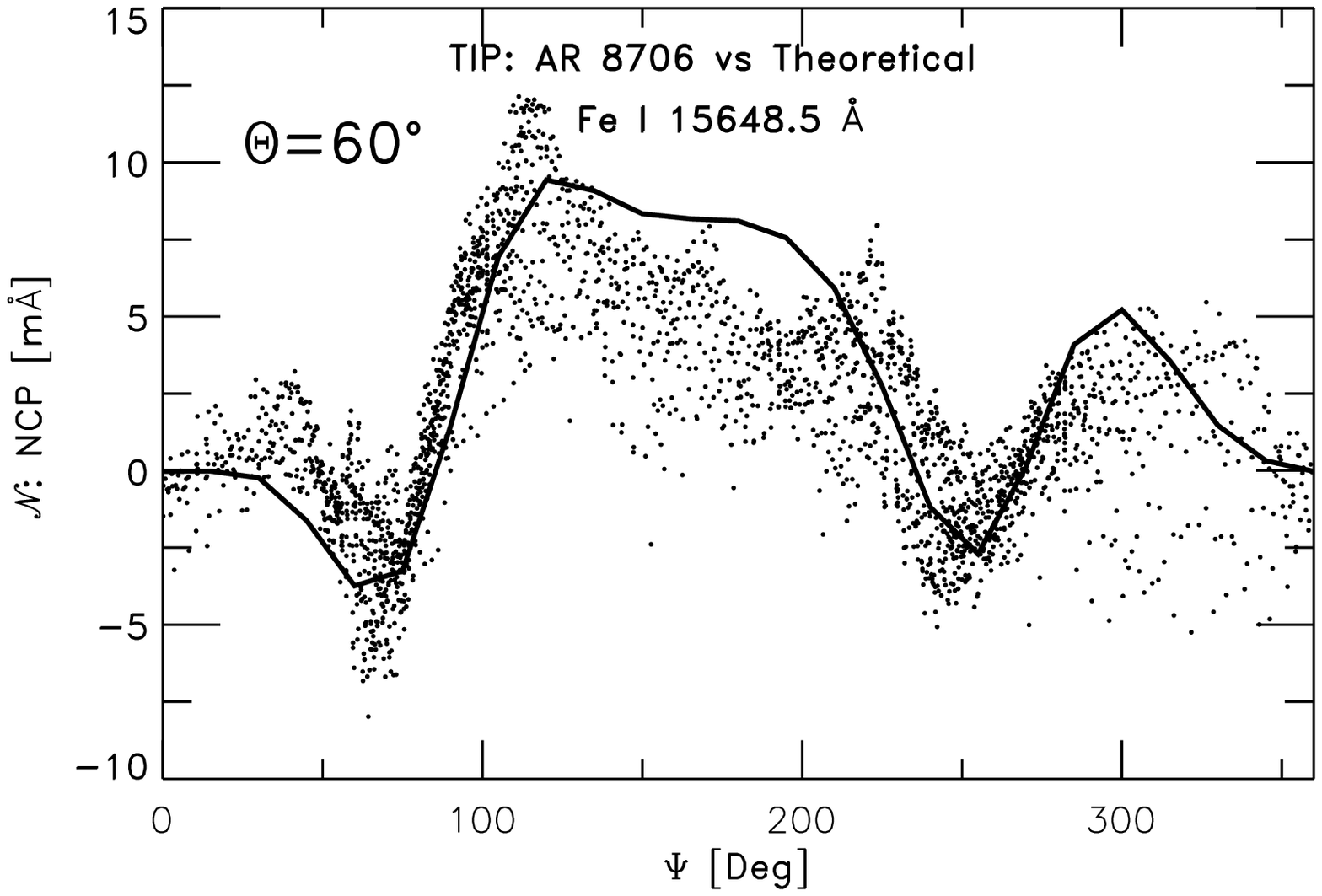}
\figcaption{Same as Figure 3 but for and Fe I 6302.5 \AA~ and $\Theta=38\deg$ (top
panel) and Fe I 15648.5 \AA~ and $\Theta=60\deg$ (bottom panel). Observations
of $\mathcal{N}(\Psi)$ for two different sunspots observed at those heliocentric
angles in these two spectral lines are displayed by the dots (same data as underlying 
Figs.~3-4 in paper II).} 
\end{center}

The magnetohydrostatic equilibrium for horizontal flux-tubes
imposes large temperatures in the tube's lower half. This yields a 
$\tau=1$ level that is always formed within the upper middle-half of the flux tube 
(see Fig.~2 in paper I), just as in our Fig.~1. Therefore, the lower half of the flux-tube
does not significantly affect the emergent radiation, so that the main 
difference between the model employed in this work and the horizontal flux-tube 
model is the addition of the convective flow (Eqs.~3 through 8).

To investigate the effect that these convective motions have on the generated
NCP, we have calculated the NCP produced by individual rays cutting through
a penumbral filament (in the same way as in Figure~2) but switching off the
Evershed effect: $V_e = 0$. An example is presented in Figure 5
for a filament located at disk center ($\Theta=0\deg$) and at the line-of-symmetry
of the sunspot ($\Psi=0^{\circ}$). We have carried out the experiment for two different
convective velocities: $V_c = 1$ (solid lines) and $3$\kms(dashed lines).
Note that, in this particular example the results would have been the same
even if a horizontal Evershed flow was present, $V_e \ne 0$ (Eq.~4), since it
does not contribute to the LOS-velocity.

According to Figure 5, the amount of NCP does not exactly scale linearly with $V_c$. This is due to the fact
that $V_c$ does not necessarily represent the convective velocities seen in
spectropolarimetric observations, but rather the strength of the convective upflow at the filament's
center (see discussion in Section 2), which is partly hidden below the $\tau=1$ level (see Fig.~1). 
For $V_c = 1$\kms~ the generated NCP is always smaller than 3 m\AA~ (absolute value). The results for the 
Fe I 6302.5 \AA~ line show that, at the center of the filament (where the upflow is present) the NCP is negative, but 
it turns positive closer to the filament's edge (at the downflow lanes).

Hinode/SP (Ichimoto et al. 2008) has not so far provided a
 clear correlation between convective velocities and NCP in sunspots penumbrae close to 
disk center in the Fe I 6302.5 \AA~ line, probably due to the limited spatial resolution 
of the observations, which smears out the NCP variation across the filament. 
For Fe I 6302.5 \AA~ this makes the effect of the convective
velocity field negligible, since the NCP generated by the upflow cancels out with the NCP generated by 
the downflowing lanes. Due to the $\Delta\phi$-mechanism (M\"uller et al. 2002) Fe I 15648.5 \AA~ does not show 
such a correlation between upflows/downflows and NCP, however, we can see in Fig.~5 
that the regions with positive net circular polarization are roughly equal to the regions 
with negative one. Therefore the spatially averaged NCP also tends to cancel out in this near-infrared
spectral line. It is also worth noticing, in Figures 2 and 5, that the curves N$_{\rm m}(x)$ are not symmetric even though
the filament is located at disk center. The asymmetry is again due to the $\Delta\phi$-mechanism which is affects
more the infrared lines (red curves) than the visible lines (blue curves).

\begin{center}
\includegraphics[width=8cm]{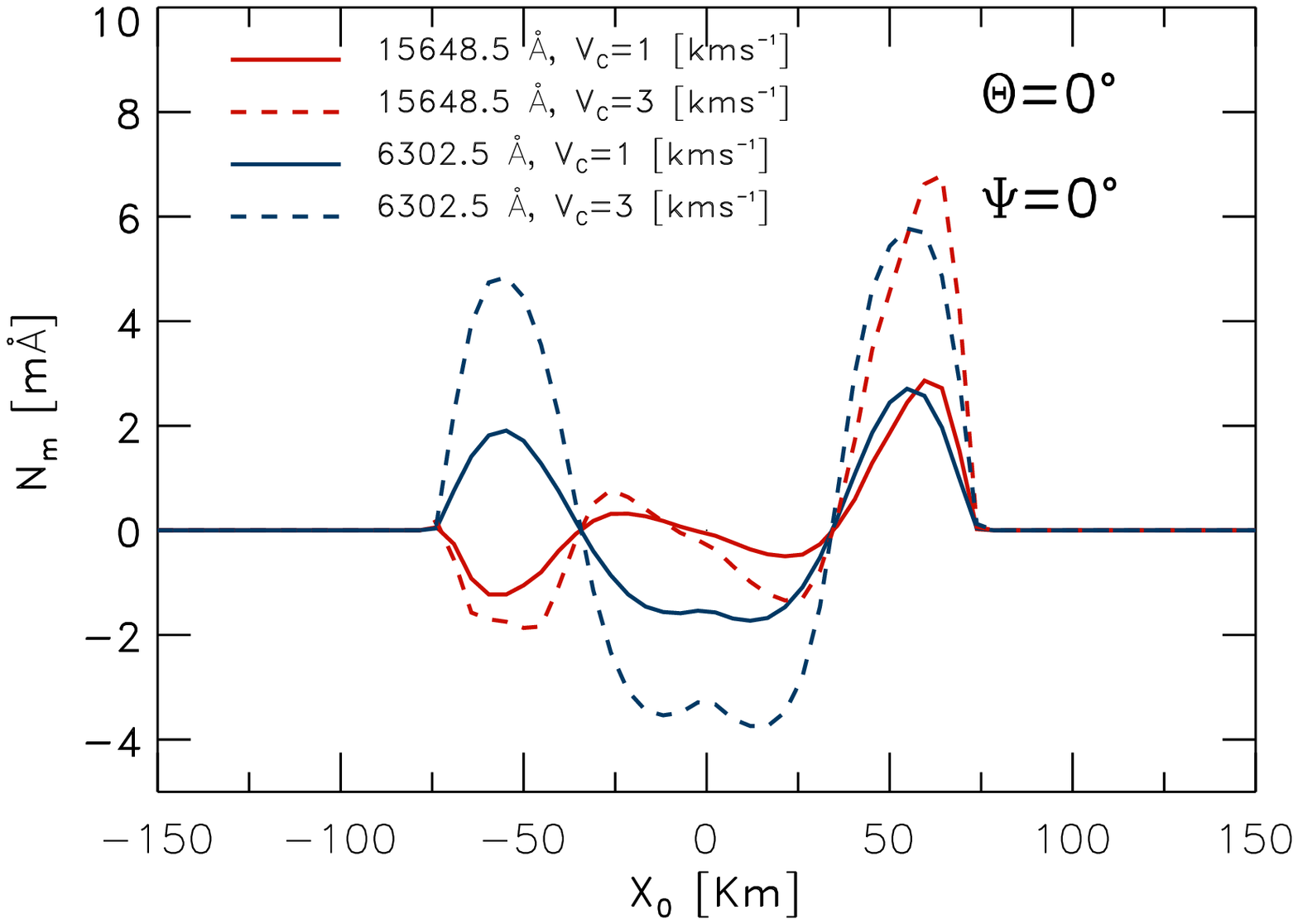}
\figcaption{Net circular polarization produced by different ray-paths, $\rm N_{\rm m}$,
cutting a penumbral filament at different $x_0$'s. This test was performed with the
following model parameters: $B_0=B_{\rm f0}=1000$ G, $\gamma_0=60\deg$, $R=75$ km, $V_e=0$\kms, 
$\Theta=\Psi=0\deg$, $V_c=1$\kms (solid lines) or $V_c=3$\kms (dashed).}
\end{center}

In Figure 6 we present different $\mathcal{N}(\Psi)$-curves for sunspots located
at different heliocentric angles. The first thing one realizes is that for $\Theta=0\deg$ a flat curve is obtained. 
This was to be expected because at disk center it does not matter where the filament is located 
within the sunspot (Eq.~15). In addition, the total NCP is very small ($|\mathcal{N}(\Psi)|<1$ m\AA).
This is in agreement with our previous discussion, and is due to the fact that the upflow at the 
filament's center produces an NCP opposite in sign as the downflowing lanes at its edges, 
yielding very small values once we calculate the spatial average.

\begin{center}
\includegraphics[width=8cm]{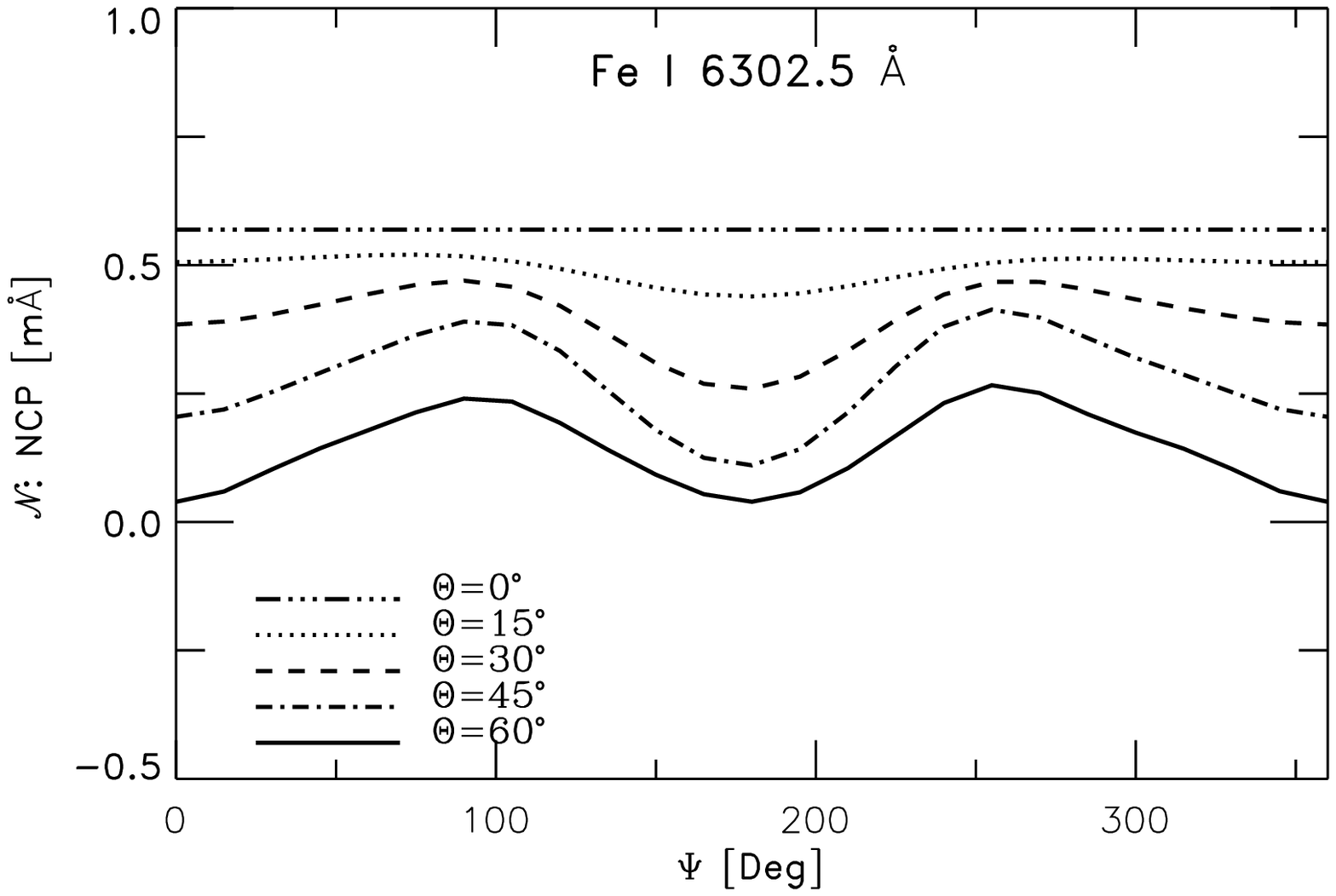} \\
\includegraphics[width=8cm]{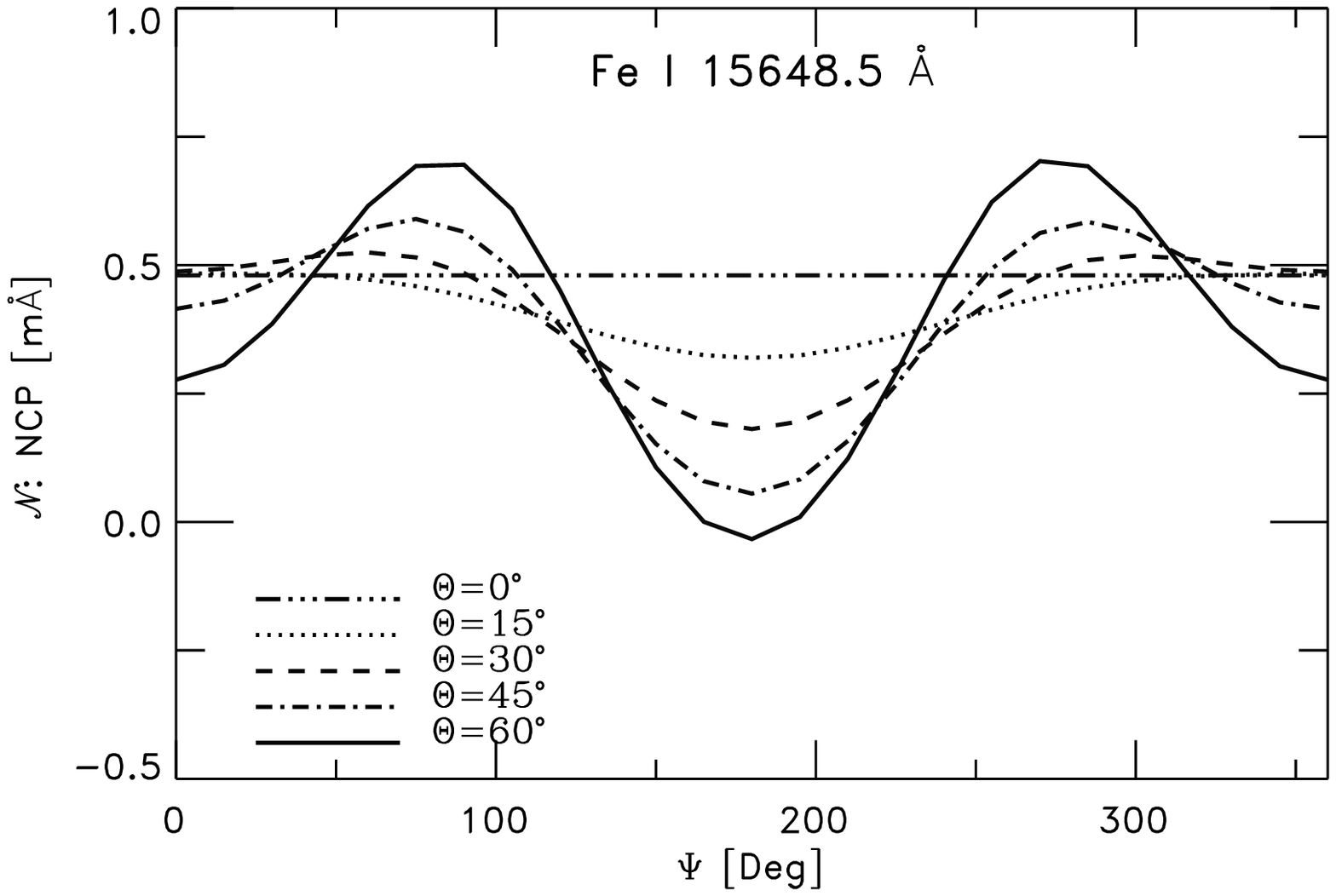}
\figcaption{Azimuthal variation of the NCP $\mathcal{N}(\Psi)$ predicted 
at different heliocentric angles $\Theta$, for a penumbral filament harboring 
no Evershed flow ($V_e = 0$) and a convective flow $V_c = 3$\kms. The rest of
the model parameters are the same as in Fig.~4: $B_0=B_{\rm f0}=1000$ G, $\gamma_0=60\deg$,
$R = 75$ km. Top panel shows the NCP calculated for Fe I 6302.5 \AA~ and the bottom
panel for Fe I 15648.5 \AA.}
\end{center}

The NCP decreases towards the limb along the line-of-symmetry of the spot ($\Psi=0,\pi$). This is 
because the projection of the convective velocity field along the observer's line-of-sight decreases,
and therefore we expect the NCP at this azimuthal position to decrease with increasing $\Theta$. Note
that this is not necessarily the case for regions perpendicular to the line-of-symmetry ($\Psi=\pi/2,3\pi/2$)
since the overturning upflow would become aligned with the observer. Indeed we observe that, for Fe I 6302.5 \AA~
(Fig.~6; top panel), at $\Psi=\pi$ and $\Psi=3\pi/2$ the NCP decreases by a smaller amount with $\Theta$ than
at $\Psi=0,\pi$. In the case of Fe I 15648.5 \AA~ (Fig.~6; bottom panel), the additional contribution of
the $\Delta\phi$-mechanism produces an increase in the NCP, perpendicular to the line-of-symmetry, as $\Theta$
increases.

We stress that in these experiments we neglected the contribution of the Evershed flow ($V_e = 0$).
If we had included it, its effect would have become larger with increasing $\Theta$, making the effect of $V_c$
even more negligible by comparison. Consequently the NCP generated by convective velocities inside penumbral
filaments is easily masked by the lack of spatial resolution, projection effects, and the additional effect
of the Evershed flow.

Finally it is important to bear in mind that the model presented here does
not transport any net energy since the temperature in the downflowing lanes is the same as in the central 
upflow\footnote{Equations 13 and 14 show that neither the density nor gas pressure, and thus also not 
the temperature, depend on the x-coordinate.}. To test what would happen in a more realistic situation where 
real convection would be present we have repeated our experiments in this section (Figs.~5 and 6) but
artificially increasing the temperature in the upflowing lane according to:

\begin{eqnarray}
\Delta T(x,z) = \Delta T_0 \left(1-\frac{z}{\sqrt{R^2-x^2}}\right) \;\;\text{if} \;\;
\begin{cases} V_z(x,z) < 0 \\ \sqrt{x^2+z^2} < R \end{cases} \;,
\end{eqnarray}

\noindent where $\Delta T_0 = 3000$ K. This value for $\Delta T_0$ has been chosen such that upflows 
provides sufficient energy to explain a penumbral brightness that is about 70 \% of the quiet Sun. 
Note that Equation 18 only applies to upflows inside the filament: $V_z < 0$ and $r=\sqrt{x^2+z^2} < R$. 
Equation 18 shows that the temperature difference vanishes at the filament's edge, where convective-like 
motions are no longer present. Under this new configuration, the results show that the actual shape for the NCP-curve
(Figs.~5-6) in Fe I 6302.5 \AA~ does not change, whereas for Fe I 15648.5 \AA~ does. On the one hand, these changes
are at the level of $\sim$ 1 m\AA, which supports our previous claim that the thermodynamic structure
plays only a secondary role in the generation of NCP. On the other hand, after modifying the temperature
in the upflow, density and gas pressure have not been modifyed in a way that is consistent with
 the equilibrium of the filament (Eq.~13-14), therefore these claims need further 
investigation with a model that allows for these differences self-consistently.

\section{Effect of the Filament's magnetic field strength on the NCP}%

Another model with distinct similarities to the structure we have studied here is
the {\it gappy penumbral model} (Spruit \& Scharmer 2006; Scharmer \& Spruit 2006),
which postulates that the penumbral filaments are formed by field-free gaps that
penetrate the penumbral magnetic field from below. Inside such field-free gaps overturning
convective motions occur. This is an advantadge, as compared to horizonal flux-tube
models (Solanki \& Montavon 1993), since convective motions are able to carry enough energy
to heat the penumbra, which in turn could explain its enhanced brightness relative to the
umbra (cf. Schlichenmaier \& Solanki 2003; Ruiz Cobo \& Bellot Rubio 2008). In the context
of the gappy penumbral model, the Evershed flow would be produced by the deflection of these
convective motions along the inclined field lines above the gap (Schamer et al. 2008b),
although it has not yet been shown that in this model the Evershed flow would be restricted
to material threaded by a magnetic field (Solanki et al. 1994).

In the previous examples (Sects.~3 and 4) we have assumed that 
the magnetic field inside the penumbral filament is as strong as 
the external field far away from the filament ($B_{\rm f0}=B_0=1000$ G). However, our model for penumbral 
filaments would present a very similar configuration, both in the magnetic 
field and the velocity field, to the gappy penumbral model if we set the 
field strength inside the filament to zero: $B_{\rm f0}=0$. Very recently, however,
Scharmer (2008) has acknowledged the possibility of a non-zero (although strongly reduced) 
magnetic field inside the \emph{field-free gap} (cf. Brummell et al. 2008; Rempel et al. 2009).

The azimuthal variation of the NCP, $\mathcal{N}(\Psi)$, for a penumbral filament observed 
away from disk center for various field strengths inside the filament is presented in Fig.~7
for Fe I 6302.5 (top panel, for $\Theta=38\deg$) and Fe I 15648.5 \AA~ (bottom panel, for $\Theta=60\deg$). 
The case of $B_{\rm f0}=1000$ G is indicated by solid lines in this figure, which are identical to
the solid lines in Fig.~4, which reproduce very well the observations. However, when the field strength inside 
the filament drops below 1000 G the discrepancy between theoretical and observed curves
becomes clear. Similar discrepancies appear also at other heliocentric angles.

In particular, for $B_{\rm f0} < 500$ G, the NCP produced
by an almost field-free filament is always negative at all azimuthal angles
in both spectral lines, and therefore does not reproduce the correct sign 
of the NCP. In addition, the multi-peak structure observed in Fe I 15648.5 
\AA~ disappears completely for $B_{\rm f0} < 500$ G, which is contrary to observations.
These computations imply a value of $B_{\rm f0}$ not much below 1000 G, which in agreement with the 
findings of Borrero \& Solanki (2008) who found that in the 
outer penumbra, the magnetic field inside penumbral filaments is not weaker than in the external field.

\begin{center}
\includegraphics[width=8cm]{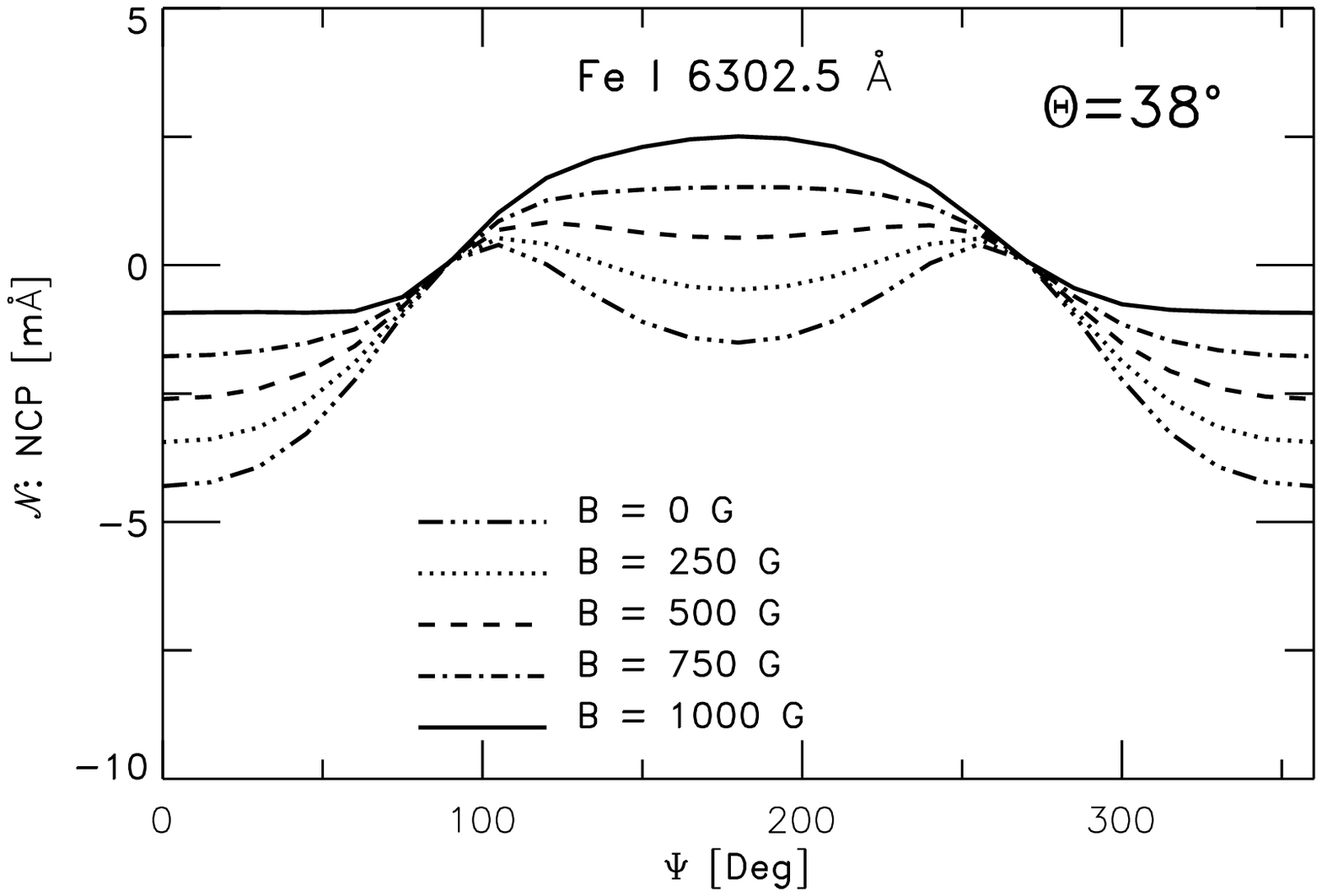} \\
\includegraphics[width=8cm]{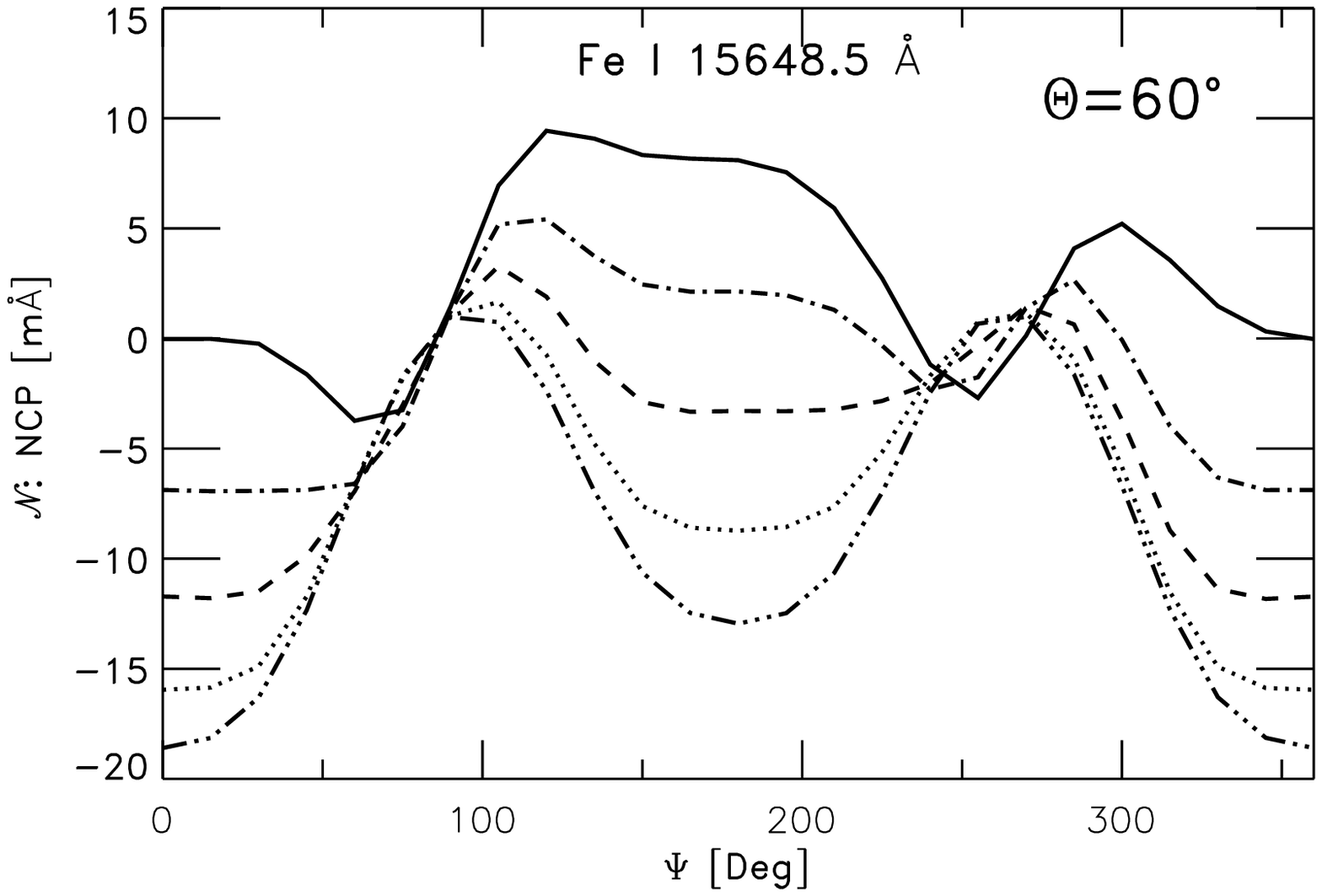}
\figcaption{Azimuthal variation of the NCP $\mathcal{N}(\Psi)$ predicted at 
$\Theta=38\deg$ for Fe I 6302.5 \AA~ (top panel), and at $\Theta=60\deg$
for Fe I 15648.5 \AA~ (bottom panel). Note that solid lines ($B_{\rm f0}=1000$ G) are the same
as in Figure 4. This case corresponds to a strong magnetic field inside the filament, and is
able to reproduce the observations satisfactorily. The other curves refer to
different values of $B_{\rm f0}$, as marked in the upper panel. The model parameters
employed here are: $B_0 = 1000$ G, $R = 75$ Km, $\gamma_0 = 60\deg$, 
$V_e = 6$\kms, $V_c = 1$\kms.}
\end{center}

\newpage
\section{Effect of other model parameters}%

In order to investigate whether our results are effected by our particular choice of
model parameters we have studied the effect of these parameters on the 
$\mathcal{N}(\Psi)$ curves. For example, the effect of the inclination of the external field. 
The idea behind this is that a smaller $\gamma_0$ increases the gradient in the inclination 
along the line-of-sight, so that a larger NCP should be generated through
 the $\Delta\gamma$-mechanism (S\'anchez Almeida \& Lites 1992). Using 
$\gamma_0 = 45^{\circ}$ indeed increases the amount of NCP (for Fe I 6305.5 \AA~ only), however it did 
not have any significant impact on the overall shape of the $\mathcal{N}(\Psi)$ curves.

We have also employed other models for the external atmosphere.
 Instead of the hot umbral model from Collados et al. (1994)
we adopted the mean penumbral model by del Toro Iniesta et al. (1994), which is about
1200 K hotter than the former at $\tau_5=1$ and possesses a steeper gradient
in temperature. Again, no significant differences were observed,
supporting our earlier statement (Sect.~2) that the thermodynamic details play only a minor role.
Of course, if the thermodynamics change dramatically noticeable differences do appear.
For example, using the cool (instead of hot) umbral model from Collados et al. (1994)
has the effect of yielding very small, $|\mathcal{N}(\Psi)|<1$ m\AA, values for the net circular polarization
in Fe I 6302.5 \AA, in clear disagreement with observations. This happens because this umbral model is very cold
and therefore the lower level of the atomic transition depopulates, which produces spectral lines that are far
from their saturation point, becoming less sensitive to the gradients along the line-of-sight
(Grossmann-Doerth et al. 1989; Borrero et al. 2004).

We have also studied the effects of other possible convective velocity fields. For
example, consider:

\begin{eqnarray}
{\bf V_{\rm f}}\deppp = V_e\ey + V_c \left\{1-\frac{r}{R}\right\}\er-V_c \cos\theta \et
\end{eqnarray}

This velocity field produces a very similar convective pattern as the one 
described in Eqs.~3-8 (see also Fig.~1), but it does not satisfy mass conservation
inside the penumbral filament. In spite of this, we have repeated most of the calculations
presented in this paper using this velocity field and found that it produces essentially
the same results as the more realistic flow that conserves mass.

Another parameters that affects the NCP-curves is the percentage of the resolution element
that is assumed to be occupied by the penumbral filament (filling factor; see footnote in Section 3).
A smaller filling factor will scale the $\mathcal{N}(\Psi)$-curves proportionally. However, on the one hand
 the model parameters we have chosen are meant to model the conditions in the middle-penumbra, where the
filling factor of the filament is seen to peak (see Bellot Rubio et al. 2004, Borrero et al. 2005). On the other
hand a decrease of a 25 \% in the filling factor can be compesated by an similar increase in the magnitude of the
Evershed flow $V_e$ or a decrease in the inclination of the external field $\gamma_0$ (that increases in the gradient in the
inclination of the magnetic field as the line-of-sight passes from the external atmosphere to the inside of the filament).

\vspace{0.9cm}
\section{Conclusions}%

We have developed a magnetohydrostatic model of a penumbral filament embedded in a
surrounding potential field. The MHS equilibrium imposes a density, pressure and temperature
structure inside the penumbral filaments such that the $\tau=1$-level is formed inside the
filament. Consequently, we do not need to specify its sub-surface structure,  which could be
in the form of a flux-tube (filament with circular cross section) or in the form
of a vertically elongated plume. Inside the filament we assume the presence of the 
Evershed flow along its axis and of a convective velocity field 
perpendicular to it. The filament's magnetic field is imposed to be homogeneous
in its interior.

By means of Stokes radiative transfer calculations, we have shown that this model
is able to reproduce the observed azimuthal variation of the net circular 
polarization $\mathcal{N}(\Psi)$, observed at different heliocentric angles for two different (visible
and near-infrared) Fe I lines.

We have also studied the effect of the convective velocity field on the generated
 $\mathcal{N}(\Psi)$-curves. We have found that its effect is much smaller
 than the NCP generated by the Evershed flow. In addition, the NCP generated by the
convective downflows ($\mathcal{N}>0$) partly cancels with the NCP generated by the upflow
at the filament's center ($\mathcal{N}<0$).

Finally, we have employed our model to study the NCP generated by field-free gaps
(Spruit \& Scharmer 2006) and have found that this model does not reproduce
satisfactorily the observed NCP. For that to happen, the magnetic field inside the
filament should be around 1000 G, which is not compatible with the concept of a 
field-free gap.

Our results do not, by themselves, rule out the field-free gap model, since the model employed here
is still rather simple, although it does account for the main features of the penumbral fine structure.
A more elaborate model based on field-free gaps could still
yield NCP curves closer to the observed ones.

In summary, our investigation confirms that the net circular polarization is produced
mainly by the Evershed flow in filaments filled with a rather strong horizontal field,
and embedded in an inclined magnetic field, as originally proposed by Solanki \& Montavon
(1993) and worked out in greater detail by Mart{\'\i}nez Pillet (2000), M\"uller et al. (2002), 
Schlichenmaier et al. (2002), Borrero et al. (2007) and others.

More elaborate models are already available thanks to recent 3D MHD simulations (Sch\"ussler \&
V\"ogler 2006, Heinemann et al. 2007, Rempel et al. 2008, 2009). These simulations, reveal a complex
picture that shares similarities and differences with both the flux-tube and the gappy penumbral model
(see Borrero 2009, Schlichenmaier 2009). In a next step it is important to introduce non-grey
radiative energy transfer into such simulations, so that similar analyses as carried out here can
be performed.

\acknowledgements{This work was partly supported by the WCU grant No. R31-10016 from the
Korean Ministry of Education, Science and Technology}

\newpage
\begin{appendix}
\section{Determination of the filament's convective flow through mass conservation}

In this section we derive a velocity field inside penumbral filaments that conforms
with mass conservation for a prescribed density and gas pressure stratification inside the filament.
Gas pressure and density have been obtained under hydrostatic equilibirum in Section 2 of this paper
and are given by Eqs.~13 and 14, respectively. In addition, in order to satisfy the boundary
conditions at the filament's boundary (Eqs.~9-10) the radial and angular component of the velocity
field must vanish at the filament's boundary: $V_{\rm fr}\depp=V_{\rm f\theta}\depp=0$. The final requirement
is that the velocity flow must be convective, that is, with an upflow at the filament's center and
downflowing lanes at its edges.\\

We start by writing down the condition: $\nn (\rho\fil {\bf V}\fil)=0$ in polar coordinates. Unless otherwise specified
we will always refer to the filament and therefore the subindex 'f' is implied throughout this section.

\begin{eqnarray}
V_r \frac{\partial \rho}{\partial r} + \frac{V_\theta}{r}\frac{\partial \rho}{\partial \theta}
+\frac{\rho}{r} \left\{\frac{\partial (rV_r)}{\partial r}+\frac{\partial V_\theta}{\partial \theta}\right\} & = & 0 \;.
\end{eqnarray}

We now rewrite A1 as:

\begin{eqnarray}
\frac{\partial V_\theta}{\partial \theta} +q\deppp V_\theta & = & m\deppp \;,
\end{eqnarray}

\noindent where $q(r,\theta)$ and $m(r,\theta)$ are:

\begin{eqnarray}
q\deppp & = & \frac{1}{\rho}\frac{\partial \rho}{\partial \theta}\;, \\
m\deppp & = & -\left\{\frac{\partial (rV_r)}{\partial r}+\frac{r V_r}{\rho}\frac{\partial \rho}{\partial r}\right\} \;.
\end{eqnarray}

Equation A2 is a first order linear partial differential equation that can be solved with the help
of an integrating factor $i\deppp$, given by:

\begin{eqnarray}
i\deppp = \exp\left\{\int q\deppp d\theta\right\} = \exp\left\{\int \frac{1}{\rho}\frac{\partial \rho}{\partial \theta} 
 d\theta\right\} = \rho\deppp \;,
\end{eqnarray}

Equation A5 shows that the integrating factor is indeed the density. Multiplying the left and right hand
sides of Eq.~A2 by the density, yields the solution for $V_\theta$ as:

\begin{eqnarray}
V_\theta \deppp = \frac{1}{\rho\deppp}\left\{\int \rho\deppp m\deppp d\theta + C(r) \right\} =
 \frac{1}{\rho\deppp}\left\{-\int \frac{\partial (r\rho V_r)}{\partial r} d\theta + C(r) \right\}
\end{eqnarray}

\noindent where $C(r)$ is an integration constant that can depend of the radial coordinate. For simplicify we will now make 
the further assumption that $V_r$ depends only on the radial distance from the filament's center: $V_{\rm fr}(r)$. 
With this, we can simplify Eq.~A6 to:

\begin{eqnarray}
V_\theta \deppp = \frac{-1}{\rho\deppp}\left\{r V_r\int \frac{\partial \rho}{\partial r}d\theta + \frac{\partial
(r V_r)}{\partial r} \int \rho d \theta - C(r) \right\}\;.
\end{eqnarray}

Now, according to Eq.~14 in Section 2 of the paper, the filament's density is given by:

\begin{eqnarray}
\rho\fil\deppp = \rho\sur(z) + r \delta \sin \theta \;,
\end{eqnarray}

 \noindent where we have only subtituted $z = r \sin \theta$ and $\delta=\frac{B_0^2\cos^2\gamma_0}{\pi g R^2}$
(Eq.~8). Now, the denstity stratification of the external atmosphere $\rho\sur(z)$ can be written in terms of the
density at $z=0$: $\rho_0$ and its density scale-height $H\sur$:

\begin{eqnarray}
\rho\sur(z) = \rho_0 e^{-z/H\sur} = \rho_0 e^{-r \sin \theta/H\sur} = \rho\sur\deppp
\end{eqnarray}

In general, the density scale-height varies with height, however, over the range of heights we are interested in: 
$z \in[0,R]$, $H_s$ can be considered to be constant. When subtituting Eq.~A9 into A8 and then into A7 we are left with two 
integrals that can be solved analitically, in terms of the hypergeometric function $_2 F_1(1/2;(1-k)/2;3/2;\cos^2\theta)$, but only
 if we perform a Taylor expansion of the density in the surrounding atmosphere $\rho\sur\deppp$.

\begin{eqnarray}
\rho\sur\deppp = \rho_0 \left[1+\sum_{k=1}^{\infty}{\frac{(-1)^kr^k}{k! H\sur^k} \sin^k\theta}\right] =
\rho_0 \left[ 1 - \frac{r}{H\sur}\sin\theta + \frac{r^2}{2H\sur^2}\sin^2\theta + \mathcal{O}(\sin^3\theta)\right]
\end{eqnarray}

Fortunately, for typical penumbral conditions we have that $R \delta> \rho_0$. In this case the 
term $r\delta\sin\theta$ in Equation A8 is the main  contributor to the filament's density $\rho\fil(z)$, which
in turn means that we can truncate the Taylor expansion of $\rho\sur(z)$ (Eq.~A10) to include only the first
two terms. In this way we can avoid dealing with hipergeometric functions and transform the integrals inside Eq.~A7 into:

\begin{eqnarray}
\int \frac{\partial \rho\fil\deppp}{\partial r} d \theta \approx \int \left[\frac{-\rho_0\sin\theta}{H\sur}\left(1-
\frac{r\sin\theta}{H\sur}\right) + \delta\sin\theta\right]d\theta = \alpha \cos \theta + \frac{r \rho_0}{2H\sur^2}
(\theta-\cos\theta\sin\theta)
\end{eqnarray}

\begin{eqnarray}
\int \rho\fil\deppp d\theta \approx \int \left[\rho_0-\rho_0 \frac{r\sin\theta}{H\sur}+r \delta \sin\theta\right]
d\theta = \rho_0 \theta + \alpha r \cos \theta
\end{eqnarray}

\noindent where $\alpha$ and $\delta$ had already been defined in Eqs.~7 and 8 in Section 2 of the paper. 
The integration constant $C(r)$ in Eq.~A7 is determined by imposing that across the center of the filament
the velocity field takes the form of an upflow: $V_{\rm f\theta}(r,\pi/2)=0$. Finally, subtituting
Eqs.~A11 and A12 into A7 yields the final result for $V_{\rm f\theta}\deppp$ given by Eq.~6. Note that Equations
6 and A7 are completely general as long as the radial component of the velocity field depends only on $r$: $V_{\rm fr}(r)$.

For the determination of $V_{\rm fr}(r)$ we are free to choose any function that vanishes at
the filament's boundary: $V_{\rm fr}\depp=0$ such that there is total pressure balance between
the filament and the surrounding atmosphere (Eq.~9). Our choice of $V_{\rm fr}(r)$ will also affect 
the functional form of $V_{\rm f\theta}\deppp$ 
(through Eq.~6). According to the discussion in Section 2 we are looking for solutions that verify: 
$V_{\rm f\theta}\depp=0$ such that the velocity term in Eq.~10 disappears. This is guaranteed if 
$V_{\rm fr}\depp=0$ (which we are already looking for) but also $\partial V_{\rm fr}\depp /\partial r=0$.
Note that our choice of $V_{\rm fr}(r)$ (Eq.~5) satisfies both conditions: both $V_{\rm fr}(r)$ 
and its derivative vanish at the filament's boundary.

\begin{eqnarray}
V_{\rm fr}(r) = V_c \left\{1-e^{-\beta \left(r-R\right)^2}\right\}
\end{eqnarray}

\end{appendix}

\end{document}